\documentclass[superscriptaddress,superbib,fleqn,prb,twocolumn]{revtex4}
\usepackage{amsfonts}
\usepackage[fleqn,reqno,tbtags]{amsmath}
\usepackage{amssymb}
\usepackage{graphicx}
\usepackage{dcolumn}
\usepackage{bibmods}

\begin{document}

\preprint{}
\title[IR of Bi2Se3]{Magneto-Infrared Study of Topological Insulator Bi$_{2}$%
Se$_{3}$}
\author{W. Yu}
\affiliation{School of Physics, Georgia Institute of Technology, Atlanta, Georgia, 30332}
\author{X. Chen}
\affiliation{School of Physics, Georgia Institute of Technology, Atlanta, Georgia, 30332}
\author{Z. Jiang}
\affiliation{School of Physics, Georgia Institute of Technology, Atlanta, Georgia, 30332}
\author{I. Miotkowski}
\affiliation{Department of Physics, Purdue University, West Lafayette, Indiana 47907}
\author{H. Cao}
\affiliation{Department of Physics, Purdue University, West Lafayette, Indiana 47907}
\author{Y. P. Chen}
\affiliation{Department of Physics, Purdue University, West Lafayette, Indiana 47907}
\author{D. Smirnov}
\affiliation{National High Magnetic Field Laboratory, Tallahassee, Florida 32310}
\author{L.-C. Tung}
\affiliation{Department of Physics and Astrophysics, University of North Dakota, Grand
Forks, North Dakota 58202}
\affiliation{National High Magnetic Field Laboratory, Tallahassee, Florida 32310}
\date{\today }

\begin{abstract}
We present a magneto-infrared spectroscopic study of thin $\text{Bi}_{2}$$%
\text{Se}_{3}$ single crystal flakes. Magneto-infrared transmittance and
reflectance measurements are performed in the Faraday geometry at $4.2$K in
a magnetic field up to $17.5$T. Thin $\text{Bi}_{2}$$\text{Se}_{3}$\ flakes
(much less than $1\mu $m thick$)$ are stabilized on the Scotch tape, and the
reduced thickness enables us to obtain appreciable far-infrared transmission
through the highly reflective $\text{Bi}_{2}$$\text{Se}_{3}$ single
crystals. A pronounced electron-phonon coupling is manifested as a Fano
resonance at the $\alpha $ optical phonon mode in Bi$_{2}$Se$_{3}$,
resulting from the quantum interference between the optical phonon mode and
the continuum of the electronic states. However, the Fano resonance exhibits
no systematic line broadening, in contrast to the earlier observation of a
similar Fano resonance in $\text{Bi}_{2}$$\text{Se}_{3}$ using
magneto-infrared reflectance spectroscopy.
\end{abstract}

\volumeyear{year}
\volumenumber{number}
\issuenumber{number}
\eid{identifier}
\received[Received text]{date}
\revised[Revised text]{date}
\accepted[Accepted text]{date}
\published[Published text]{date}
\startpage{101}
\endpage{102}
\maketitle

Bismuth selenide (Bi$_{2}$Se$_{3}$) is well known for its large Seebeck
coefficient and thermal figure of \ merit since the 50s\cite{Iof57,Bla57},
and recently it has been shown as a physical realization\cite{Zha09,Xia09}
of \ the 3D topological insulators (TIs).\cite{Has09,Moo10,Has10,Rev} Along
with Bi$_{2}$Te$_{3}$, Sb$_{2}$Te$_{3}$ and some other II$_{2}$V$_{3}$
binary alloys, 3D topological insulators are establishing a new forefront
for the next-generation nanoelectronic, spintronic, thermoelectric and
quantum computational devices.\cite{Has09,Moo10,Has10,Rev,Plu02,Fu08} The
bulk of a TI is expected to be an insulator, because the conduction band and
the valence band are separated by a sizable bandgap. The insulating bulk is
enclosed by a robust conducting surface state, containing a single gapless
Dirac cone (for the Bi$_{2}$Se$_{3}$ case). The robust surface state is
protected by the time-reversal symmetry, and the large spin-orbit coupling
leads to several interesting spin-relavent phenomena. Such a unique band
structure has been predicted by the first principles band structure
calculation and confirmed by angle resolved photoemission spectroscopy.\cite%
{Rev} Bi$_{2}$Se$_{3}$ is usually regarded as a canonical TI, because it has
a one-valley conduction-band minimum and a one-valley valence-band maximum
occurring at the centre of the Brillouin zone separated by a sizable direct
bandgap $E_{g}\sim 0.2-0.4$eV.\cite{Mis97} Between the band edges, it spans
a single gapless Dirac cone with the Dirac point located around $0.3$eV
below the bottom edge of the conduction band.\cite{Zha09}

The unique topology of a TI leads to many interesting new coherent quantum
phenomena\cite{Qi10}, and one of them is an intrinsic magnetoelectric
coupling\cite{Tse10} that allows an applied magnetic (electric) field to
induce an effective electric (magnetic) field in the same direction. An
unusual magnetic-field dependent phonon softening observed via the Fano
resonance\cite{Fan61} at the $\alpha $ phonon mode ($\symbol{126}64$cm$^{-1}$%
) in Bi$_{2}$Se$_{3}$ has been regarded as evidence in support of the
magnetoelectric coupling in non-trivial TIs.\cite{Laf10} An applied magnetic
field along the trigonal $c$-axis induces a local electric field acting on
the Bi ions via the magnetoelectric effect, and the local electric field
causes a change in the lattice dynamics and a softening (linewidth
broadening) of the $\alpha $ phonon mode.\cite{Laf10} However, a similar
concurrent infrared reflectance study\cite{But10} and an early study\cite%
{Ric77} on $\text{Bi}_{2}$$\text{Se}_{3}$\ do not exhibit such a unique
asymmetric Fano lineshape at the $\alpha $ phonon mode. More recently, a
similar Fano resonance at the $\alpha $\ phonon mode has been observed\ in Bi%
$_{2}$Se$_{2}$Te and lightly Ca-doped $\text{Bi}_{2}$$\text{Se}_{3}$, and it
is temperature dependent with more pronounced Fano lineshape at the lower
temperatures.\cite{Pie12} In this Letter, we present a magneto-infrared
transmittance and reflectance spectroscopic study on Bi$_{2}$Se$_{3}$
ultrathin single crystal flakes at the liquid helium temperature, in which a
Fano resonance is indeed observed at the $\alpha $ phonon mode, but the
linewidth is insensitive to the increasing magnetic field. We interpret our
data, as well as that reported in previous works\cite%
{Laf10,But10,Ric77,Pie12}, using the charged phonon theory.\cite{Ric92,Cap12}

Bi$_{2}$Se$_{3}$\ is a layered compound with quintuple layers, each with
five atomic layers, stacked along the trigonal $c$ axis.\cite{Bla57} The
synthesis of single crystal $\text{Bi}_{2}$$\text{Se}_{3}$ is described in
Ref. \cite{Qi10a} and it crystallizes in a rhombohedral structure (point
group $\overline{3}mD_{3}d$) with the lattice parameters $a=4.138$\AA\ and $%
c=28.64$\AA .\cite{Book} Neighboring quintuple layers are bounded by weak
van der Waals forces, allowing one to exfoliate Bi$_{2}$Se$_{3}$ thin layers
from a larger crystal. Bi$_{2}$Se$_{3}$ single crystals of nominal thickness
are highly reflective and impermeable to infrared light. For the
magneto-infrared transmittance measurements, the ultrathin semi-transparent
Bi$_{2}$Se$_{3}$ flakes are prepared by repeatedly exfoliating layers from a
thin Bi$_{2}$Se$_{3}$ crystal laid on Scotch tape, until it becomes
permeable to infrared radiation. This process results in a sample consisting
of thousands of ultrathin flakes, and most of them exhibit a cross-sectional
area of around $100-300\mu $m$^{2}$.

The Bi$_{2}$Se$_{3}$ flakes on Scotch tape are then placed in a liquid
helium cryostat held at $4$K and subject to an applied magnetic field
parallel to the $c$-axis up to $17.5$T, i.e. in the Faraday geometry.
Magneto-infrared transmittance spectra are measured by a Fourier transform
infrared (FTIR) spectrometer using light pipe optics in the experimental
setup shown in Fig. 1 (a). Here, the light from a broadband infrared light
source (a Mercury lamp) is modulated by a Michelson interferometer to
produce a light beam with its intensity modulated according to the
difference in the lengths of the two optical arms (a FT-modulated beam).\cite%
{BookG} Next, as shown in Fig. 1 (a), the FT-modulated beam is steered by
mirrors and sent to a sample via a highly polished metallic pipe (a light
pipe). At the end of the light pipe, a parabolic cone is used to focus the
light onto a spot of several $mm$ in diameter on the sample as shown
in Fig. 1 (b). The intensity of the light transmitted through the sample is
measured by a Si chip inside a cavity (a bolometer). The light intensity is
measured by the minute change in the chip's temperature caused by absorbing
the energy of the FT-modulated beam. The minute temperature change results
in a change of the chip's resistance and thus the voltage drop $V_{B}$
across the two contacts on the Si chip as shown in Fig. 1 (c). The change of
the voltage drop, which is proportional to the light intensity received at
the bolometer, is then amplified and sent back to the FTIR spectrometer to
be Fourier transformed in order to obtain the transmittance spectrum.
Reflectance spectra are obtained using a similar configuration, though the
light reflected from the sample is collected and measured by the bolometer.

\begin{figure}[tp]
{{{%
\includegraphics[
]{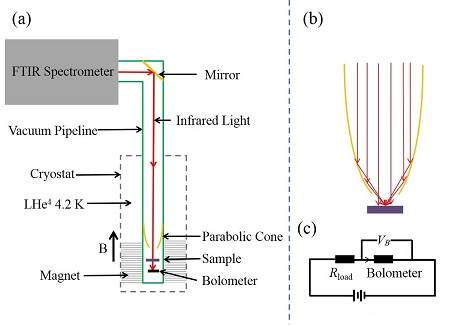}}}}
\caption{(Color online) Diagram of the experimental setup for low
temperature magneto-infrared transmittance measurements. (a) The optics
bench for steering the infrared light to a sample in a cryostat. (b) The
parabolic cone that focuses the light onto the sample. (c) The electric
circuit for the bolometer.}
\end{figure}

To extract the effect that is induced by an applied magnetic field, the
measured spectra are normalized by the reference spectra measured at zero
magnetic field. A set of normalized magneto-transmittance spectra $%
T(B)/T(B=0T)$ is plotted in Fig. 2 (a). As one can clearly see, the applied
magnetic field induces a noted change in the transmittance spectrum which
invokes the lineshape of Fano resonance at around $62$cm$^{-1}$ ($8.065$cm$%
^{-1}\simeq 1$meV). This frequency coincides with the $\alpha $ phonon mode
in Bi$_{2}$Se$_{3}$ and a weak absorption dip due to the infrared-active $%
\alpha $ phonon mode can be seen in the transmittance spectrum at zero
magnetic field. Figure 2 (a) shows the magneto-infrared transmittance
spectra, normalized to zero field. As one can see, the presence of magnetic
field induces a transfer of the optical oscillator strength from the phonon
mode to the higher frequency side of the phonon mode, thus resulting in a
rise of the normalized transmittance at the phonon frequency followed by a
broad dip on the higher frequency side. The transfer of the optical
oscillator strength increases with increasing magnetic field and the
normalized magneto-reflectance spectra collaboratively confirms the same
phenomenon as shown in Fig. 2 (b) with a dip at the phonon frequency
(decrease in oscillator strength) followed by a rise on the higher frequency
side (increase in oscillator strength) in the normalized reflectance
spectra. 
\begin{figure}[tp]
{{{%
\includegraphics[
]{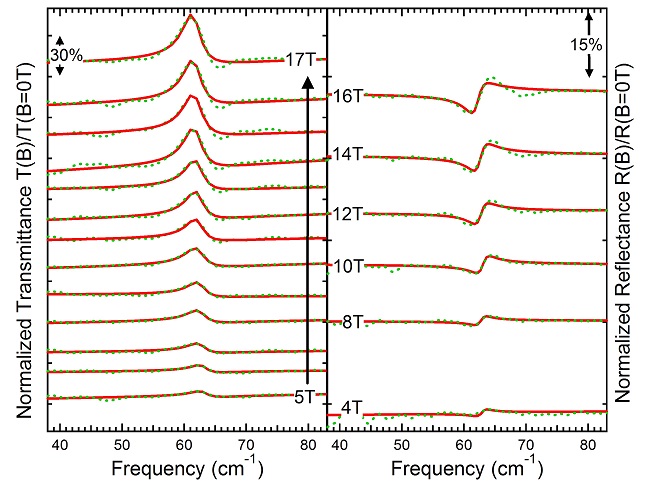}}}}
\caption{(Color online) (a) Normalized magneto-infrared transmittance and
reflectance spectra. (a) The normalized transmittance spectra $T(B)/T(B=0$T$)
$ are plotted in (green) dotted lines, obtained by the ratio to a reference
transmittance spectrum measured at $B=0$T. As a result, it reveals the
changes in transmittance induced by an applied magnetic field. (b)
Similarly, the \ normalized reflectance spectra $R(B)/R(B=0$T$)$ are plotted
in (green) dotted lines, obtained by the ratio to a reference reflectance
spectrum measured at $B=0T$. The spectra are shifted vertically for
clarity. The (red) solid lines represent the best fits to the Fano
resonance, and they are overlaid with the measured spectra for comparison.}
\end{figure}

Generally speaking, a Fano resonance results from the quantum interference
between two coupled transition pathways: one via a discrete excited state
and the other via a continuum of states.\cite{Fan61} It is ubiquitous across
several branches of physics and it results in an asymmetric scattering cross
section due to the quantum interference between the wavefunctions of the two
transition pathways. In this study, the discrete mode is the $\alpha $
optical phonon mode in $\text{Bi}_{2}$$\text{Se}_{3}$\cite{Book,Ric77},
while the continuum of states is the transitions from the lower conduction
band to the empty states in the upper conduction band. The two optical
pathways are coupled via a strong electron-phonon coupling. It was also
known that the optical phonon modes in Bi$_{2}$Se$_{3}$ and Bi$_{2}$Te$_{3}$
are stronger than expected in polar materials\cite{Ric77}, which can be a
result of rather strong electron-phonon coupling.

The observed magnetic-field induced modification of the Fano effect is
unusual for a non-magnetic system without undergoing through a phase
transition.\cite{Laf10} The Fano resonance in Bi$_{2}$Se$_{3}$ is modified
by the applied magnetic field via tuning the electronic transitions.
Specifically, the strong electron-phonon coupling has been attributed to the
magnetostriction in a system with large spin-orbit coupling and the
topological magnetoelectric effect in TIs.\cite{Laf10} An applied magnetic
field modifies the local electric field acting on Bi ions via the
magnetoelectric effect\cite{Qi08,Ess09} and thus changes the local lattice
dynamics and the optical phonon mode.

The Fano resonance induced change in optical conductivity can then be
written as $\Delta \sigma _{Fano}(\omega )=\frac{\omega _{p}^{2}}{4\pi
\gamma }\frac{q^{2}+2qz-1}{q^{2}(1+z^{2})}$, where $z=\frac{\omega -\omega
_{\alpha }}{\gamma }$.\cite{Fan61,Laf10} The square of the plasma frequency $%
\omega _{p}$ is proportional to the optical strength of the Fano mode, $%
\omega _{\alpha }$ is the $\alpha $\ phonon energy, $\gamma $ is the
linewidth, and $q$ is the dimensionless Fano parameter which characterizes
the resonance lineshape. When $\left\vert q\right\vert \ll 1$, the Fano
effect results in an anti-resonance (enhanced transmittance), and it becomes
a resonance (absorption) when $\left\vert q\right\vert \gg 1$. The sign of
the Fano parameter describes the direction (toward the higher or the lower
frequency side) of the optical strength transfer and its magnitude reflects
the degree of the coupling between the discrete state and the continuum of
the electronic states. The effect on the normalized
transmittance/reflectance spectra can be described quantitatively as the
following,%
\begin{equation}
\text{Absorption }A=A_{0}\frac{q^{2}+2qz-1}{q^{2}(1+z^{2})},
\end{equation}%
where $A_{0}$ stands for the amplitude of the Fano resonance in terms of the
absorption rate normalized by the zero-field reference spectrum. The (red)
solid lines in Fig. 2 show the best fits to the data using Eq. (1). Here,
each pair of the normalized transmittance and reflectance spectra are fitted
by the same set of parameters.

Figure 3 summarizes the fitting parameters obtained using Eq. (1) as a
function of the magnetic field. First, as one can see, the phonon energy
slightly decreases with increasing magnetic field. The phenomenon is
different from that reported in Ref. \cite{Laf10} where the phonon energy is
practically a constant. In addition, in Fig. 3(b), the linewidth $\gamma $
stays roughly a constant with increasing magnetic field, whereas the earlier
study in Ref. \cite{Laf10} reported a substantial broadening of the
linewidth (two-fold increase from zero field to 8T). In both studies, the
optical strength are transferred away from the optical phonon mode with
increasing magnetic field. It is transferred to the higher frequency side
with positive $q$ values in Fig. 3 (c), whereas the $q$ values are negative
in Ref. \cite{Laf10}. The amplitude of the Fano resonance increases with
increasing magnetic field, as shown in Fig. 3 (d). 
\begin{figure}[tp]
{{{%
\includegraphics[
]{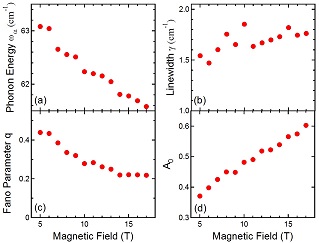}}}}
\caption{(Color online) (a) The energy of the optical phonon mode $\protect%
\alpha $ as a function of the magnetic field. (b) The linewidth of the Fano
resonance $\protect\gamma $\ as a function of the magnetic field. (c) The
dimensionless Fano parameter $q$ as a function of the magnetic field. (d)
The amplitude $A_{0}$ as a function of the magnetic field.}
\end{figure}

To explain the lineshape discrepancies reported in TI systems\cite%
{Laf10,But10,Ric77,Pie12}, we examine the possible origins of the observed
Fano resonance. We note that the magnetic-field tunable Fano resonance
around the optical phonon mode has been observed in several graphitic
systems and it can be explained by the "charged-phonon theory".\cite%
{Ric92,Cap12} Specifically, depending on the relative energies of the
optical phonon mode and the continuum of the electronic states, the optical
phonon mode can "\textit{borrow}" the optical strength (charges) from the
electronic states and vice versa. The Fano resonance observed in this work
can be qualitatively explained if the continuum of the electronic transition
is assigned to the transitions from the Fermi level in the lower conduction
band to the empty states in the upper conduction band (see Ref. \cite{Kul99}
for the band structure of Bi$_{2}$Se$_{3}$). When the Fermi level moves to
lower energy, the energy of the continuous transitions shifts to higher
energy. When the bottom of the continuum is much higher than the phonon
energy, the lineshape remains Lorentzian (i.e. $\left\vert q\right\vert \gg
1 $) with enhanced optical strength borrowed from the electronic
transitions, which explains why the Fano lineshape was not observed in\ low
density (Fermi energy) samples.\cite{Ric77,But10} When the Fermi energy is
sufficiently high, the continuum of transitions shifts to lower energy and
the asymmetric Fano lineshape is observed.\cite{Laf10,Pie12} With increasing
magnetic field, the diamagnetic shift of the Fermi level modifies the
continuum of transitions toward lower energy, which enhances the optical
strength transfer from the optical phonon mode to the electronic
transitions, accompanied by smaller $\left\vert q\right\vert $ values. The
linewidth and the direction of the transfer will depend on the density of
states of the continuum, the imperfection of the lattice, and the type of
the interaction (i.e., attractive or repulsive) between the optical phonon
mode and the continuum. Positive $q$ values may indicate that the density of
states is larger on the higher frequency side, and the negative $q$ values
indicate the opposite. The linewidth is relatively large and insensitive to
the applied magnetic field in this study, indicating that it is dominated by
the lattice imperfection of the flakes; whereas the linewidth increases with
increasing magnetic field in the earlier study\cite{Laf10}, because it was
carried out on a large $\text{Bi}_{2}$$\text{Se}_{3}$\ single crystal.

Finally, we note that most of the recent magneto-optical studies on TIs
focus on the dynamics of the charge carriers via either conductivity
analysis and/or the cyclotron resonance spectroscopy.\cite%
{Laf10,But10,Pie12,Ste07,Sch12,Cha14,Orl15} Curiously, we did not observe
any absorptions that can be attributed to the cyclotron resonance below $600$%
cm$^{-1}$ and the recently reported interband Landau level transitions
reside in the frequency range of several thousands cm$^{-1}$.\cite{Orl15}

In summary, we studied the magnetic-field tunable Fano resonance occurred at
the $\alpha $ phonon mode of Bi$_{2}$Se$_{3}$ via magneto-infrared
transmittance and reflectance spectroscopy. The observed Fano resonance can
be attributed to the electron-phonon coupling between the $\alpha $ optical
phonon mode and the continuum of the electronic transitions from the Fermi
level in the lower conduction band to the empty states in the upper
conduction band. The behavior of this Fano resonance, as well as the
discrepancies in previous studies, can be understood within the framework of
the charged phonon theory. The origin of the strong electron-phonon coupling
may be a result of the topological magnetoelectric effect, though further
proofs are required.

\begin{acknowledgments}
This work is supported by the ND-EPSCoR (EPS-0814442), by the University of
North Dakota, and by the state of North Dakota. TI crystal synthesis at
Purdue University is supported by the DARPA MESO program (Grant
N66001-11-1-4107). The IR measurement is supported by the DOE
(DE-FG02-07ER46451) and carried outw at the National High Magnetic Field
Laboratory, which is supported by NSF Cooperative Agreement No. DMR-0654118,
by the State of Florida, and by the DOE.
\end{acknowledgments}


\begin{thebibliography}{99}
\bibitem{Iof57} A.F. Ioffe, \textit{Semiconductor Thermoelements and
Thermoelectric Cooling} (London: Infosearch 1957)

\bibitem{Bla57} J. Black, E. M. Conwell, L. Seigle and C. W. Spencer,
Journal of Physics and Chemistry of Solids \textbf{2} (3), 240-251 (1957).

\bibitem{Zha09} H. Zhang, C.-X. Liu, X.-L. Qi, Xi Dai, Z. Fang and S.-C.
Zhang, Nat. Phys. \textbf{5}, 438 (2009)

\bibitem{Xia09} Y. Xia, D. Qian, D. Hsieh, L. Wray, A. Pal, H. Lin, A.
Bansil, D. Grauer, Y. S. Hor, R. J. Cava and M. Z. Hasan, Nat. Phys.\textbf{%
\ 5}, 398 (2009)

\bibitem{Has09} M. Z. Hasan, H. Lin and A. Bansil,\ Physics \textbf{2}, 108
(2009)

\bibitem{Moo10} J. E. Moore, \textquotedblleft The Birth of Topological
Insulators\textquotedblright , Nature \textbf{464}, 194 (2010)

\bibitem{Has10} M.Z. Hasan and C. L. Kane, Rev. Mod. Phys. \textbf{82}, 3045
(2010)

\bibitem{Rev} For recent reviews, see for example, M. Hasan and C. L. Kane,
Rev. Mod. Phys. \textbf{82}, 3045 (2010) and X. L. Qi and S. C. Zhang, Rev.
Mod. Phys. \textbf{83}, 1057 (2011).

\bibitem{Plu02} K. J. Pluciski, W. Gruhn, I. V. Kityk, W. Imioek, H.
Kaddouri and S. Benet, Optics Communications \textbf{204} (1-6), 355-361
(2002).

\bibitem{Fu08} L. Fu and C. L. Kane, Physical Review Letters \textbf{100}
(9), 096407 (2008).

\bibitem{Mis97} S.K. Mishra, S. Satpathy, and O. Jepsen, J. Phys.: Condens.
Matter \textbf{9}, 461 (1997)

\bibitem{Qi10} Xiao-Liang Qi, Taylor L. Hughes, and Shou-Cheng Zhang, Phys.
Rev. B \textbf{78}, 195424 (2008)

\bibitem{Tse10} W-K. Tse and A. H. MacDonald, Phys. Rev. B \textbf{82},
161104 (2010)

\bibitem{Fan61} U. Fano, Phys. Rev. \textbf{124}, 1866 (1961)

\bibitem{Laf10} A.D. LaForge, A. Frenzel, B. C. Pursley, T. Lin, X. Liu, J.
Shi and D.N. Basov, Phys. Rev. B \textbf{81}, 125120 (2010)

\bibitem{But10} N.P. Butch, K. Kirshenbaum, P. Syers, A.B. Sushkov, G. S.
Jenkins, H.D. Drew, and J. Paglione, Phys. Rev. B \textbf{81}, 241301(R)
(2010)

\bibitem{Ric77} C. R. Richter, W. Kohler, and H. Becker, Phys. stat. sol. b 
\textbf{84}, 619 (1977)

\bibitem{Pie12} P.Di Pietro, F.M. Vitucci, D. Nicoletti, L. Baldassarre, P.
Calvani, R. Cava, Y.S. Hor, U. Schade, and S. Lupi, Phys. Rev. B \textbf{86,}
045439 (2012)

\bibitem{Ric92} M.J. Rice and H.-Y Choi, Phys. Rev. B \textbf{45}, 10173
(1992)

\bibitem{Cap12} E. Cappelluti, L. Benfatto, M. Manzardo and A.B. Kuzemenko,
Phys. Rev. B \textbf{86}, 115439 (2012)

\bibitem{Qi10a} J. Qi, X. Chen, W. Yu, P. Cadden-Zimansky, D. Smirnov, N.H.
Tolk, I. Miotkowski, H. Cao, Y.P. Chen, Y. Wu, S. Qiao, Z. Jiang, Appl.
Phys. Lett. \textbf{97}, 182102 (2010)

\bibitem{Book} "Bismuth telluride (Bi$_{2}$Te$_{3}$) effective masses", 
\textit{Non-Terrahedrally Bonded Elements and Binary Compounds I}" of the
series \textit{Landolt-Bornstein-Group III Condendensed Matter} Volume 41C
by Springer Berlin Heidelberg (1998)

\bibitem{BookG} Peter R. Griffiths and James a. de Haseth, "\textit{Fourier
Transform Infrared spectrometry}" published by Johnley Wiley\& Son, Inc.
(1986)

\bibitem{Qi08} Xiao-Liang Qi, T.L. Hughes, and Shou-Cheng Zhang, Phys. Rev.
B \textbf{78}, 195424 (2008)

\bibitem{Ess09} A.M. Essin, J.E. Moore, and D. Vanderbilt, Phys. Rev. Lett. 
\textbf{102}, 146805 (2009)

\bibitem{Kul99} V.A. Kulbachinskii, N. Miura, H. Arimoto, T. Ikaida, P.
Lostak, and C. Drasar, J. Phys. Jpn. \textbf{68}, 3328 (1999)

\bibitem{Ste07} N.P. Stepanov, S.A. Nemov, M.K. Zhitinskaya, and T.E.
Svechinikova, Semiconductor \textbf{41}, 786 (2007)

\bibitem{Sch12} A.A. Schafgans, K.W. Post, A.A. Taskin, Y. Ando, X.-L. Qi,
B.C. Chapler, and D.N. Basov, Phys. Rev. B \textbf{85}, 195440 (2012)

\bibitem{Cha14} B.C. Chapler, K.W. Post, A. R. Richardella, J.S. Lee, J.
Tao, N. Samarth, and D.N. Basov, arXiv:1405.4916v1 [cond-mat.mtrl-sci] (2014)

\bibitem{Orl15} M. Orlita, B.A. Piot, G. Martinez, N.K. Sampath Kumar, C.
Faugeras, M. Potemski, C. michael, E. M. Hankiewicz, T. Brauner, C. Drasar,
S. Schreyeck, S. Grauer, K. Brunner, C. Gould, C. Brune, and L. W.
Molenkamp, Phys. Rev. Lett. \textbf{114}, 186401 (2015)
\end{thebibliography}
\end{document}